\documentclass[11pt,preprint]{aastex}

\shorttitle{Disk/Jet Simulations}
\shortauthors{De Villiers et al.}
\usepackage{graphicx}

\begin{document}

\title{GRMHD Simulations of Disk/Jet Systems: Application to 
the Inner Engines of
Collapsars}


\author{Jean-Pierre De Villiers, Jan Staff, Rachid Ouyed}
\affil{Dept. of Physics and Astronomy\\
University of Calgary\\
2500 University Drive NW\\
Calgary, Alberta, T2N 1N4}

\email{vjpde@ucalgary.ca; jstaff@capca.ucalgary.ca; ouyed@phas.ucalgary.ca}

\begin{abstract}
We have carried out 2D and 3D general relativistic magnetohydrodynamic 
simulations of jets launched self-consistently from accretion disks orbiting Kerr black holes and applied the results to the inner engine of the collapsar 
model of gamma-ray bursts.
The accretion flow launches energetic jets in the axial funnel region of the 
disk/jet system, as well as a substantial coronal wind. The jets feature 
knot-like structures of extremely hot, ultra-relativistic gas; the gas in these 
knots begins at moderate velocities near the inner engine, and is accelerated
to ultra-relativistic velocities (Lorentz factors of 50, and higher) by 
the Lorentz force in the axial funnel. The increase in jet 
velocity takes place in an acceleration zone extending to at least a few hundred 
gravitational radii from the inner engine. The overall energetics of the jets 
are strongly spin-dependent, with high-spin black holes producing the highest 
energy and mass fluxes. In addition, with high-spin black holes, the 
ultra-relativistic outflow is cylindrically collimated within a few hundred 
gravitational radii of the black hole, whereas in the zero-spin case the jet 
retains a constant opening angle of approximately 16 degrees. The simulations 
also show that the coronal wind, though considerably slower and colder than the 
jets, also carries a significant amount of mass and energy.  When simulation 
data is scaled to the physical dimensions of a collapsar the jets operate for a 
period ranging from 0.1 to 1.4 seconds, until the accretion disk is depleted, delivering $10^{48}$ to $10^{49}$ erg. Longer duration and higher
energies are inferred for more massive disks.
However, the presence of a strong coronal wind has important implications for 
any remnant stellar fallback material that has not migrated to the main 
accretion disk before the wind is established; our simulations suggest 
that such material would be blown out by the coronal wind. 
\end{abstract}


\keywords{Black holes - magnetohydrodynamics - jets:accretion - gamma-ray bursts}

\section{Introduction}

The numerical study of black hole accretion has been an area of ongoing 
research, beginning with the pioneering studies of Wilson (1972). Simulations 
of accretion mediated by the Magneto-Rotational Instability (MRI; Balbus \& 
Hawley, 1998) also have a long history, with various approximations used to 
mimic the properties of the inner engine (e.g., the pseudo-Newtonian 
simulations of Hawley \& Krolik, 2001). More recently, codes such as the GRMHD 
code of De Villiers \& Hawley (2003; hereafter DH03) have been used to study
accretion disks orbiting Kerr black holes and the jets that arise 
self-consistently from the accretion flow (De Villiers, Hawley \& Krolik, 2003; 
hereafter DHK). The key result from these simulations is that GRMHD is 
required to correctly capture the effect of black hole spin on the energy of
the jets (De Villiers et al., 2005; hereafter D05), which are powered by 
an MHD interaction between the accretion flow and the spinning black hole. 

In this paper, we analyze in greater detail the jets that arise in accretion 
simulations. We use a new version of the GRMHD code which features an expanded 
dynamic range, to better handle the large density contrast between disk and jet, 
as well as a significantly higher upper bound on the Lorentz factor ($W_{\rm 
max}=$ 50, instead of 8 in the older code, as discussed in the Appendix). 
Our 2D and 3D simulations follow the 
evolution of thick accretion tori located close to the marginally stable orbit, 
$r_{\rm ms}$, of the black hole. This choice of initial conditions is meant to 
establish whether the basic dynamics of the disk/jet system evolved by the GRMHD 
code can reproduce the essential features invoked 
for the inner engine in
the collapsar model 
of gamma-ray bursts (GRBs; Woosley, 1993, hereafter W93).
In this scenario, that of failed Type Ib supernovae, the core of Wolf-Rayet star has undergone 
collapse to a black hole and a thick accretion disk has formed near the 
marginally stable orbit of the black hole. The accretion flow is thought to 
power energetic jets along the rotational axis which are responsible for the 
release of brief, intense bursts of 
radiation once the jets reach the optically thin region at the surface of the stellar remnant. 
Collapsar simulations using a variety of techniques have been reported in the 
literature: 
hydrodynamic simulations, e.g. MacFadyen \& Woosley (1999); pseudo-Newtonian 
magnetohydronamic simulations, e.g. Proga et al. (2003), which improve upon 
hydrodynamic models by better representing the turbulent transport of angular 
momentum; and more recent attempts to apply GRMHD to the problem, e.g. Mizuno et 
al. (2003). The simulations reported here consider the dynamics of a long-lived 
disk/jet system in a self-consistent manner using full GRMHD, allowing us to 
better quantify the energetics of the system and the role played by black hole 
spin in the energy of the jets. 

Given the ability to simulate the relativistic dynamics of 
the inner engine of 
a collapsar, we seek to answer three questions: what is the range of 
Lorentz factors in the jets; how much energy is transported by these jets (and for how long); and how variable is the flux of energy. The answer to these
questions should help to establish whether the dynamics of the GRMHD interaction 
between disk, black hole, and jet 
in our numerical setup 
support the basic tenets of the collapsar 
model. Since the GRMHD code is scale-free, we will discuss the scaling of our 
results to the collapsar model in a separate section. We follow the notation 
used in DH03; a summary of the equations evolved by the GRMHD code and the set 
of dynamical variables available for analysis is given in the appendix, where
a discussion of the numerical determination of the Lorentz factor is also
given.

We report all simulation results in geometrodynamic units (Misner, Thorne \& 
Wheeler, 1973), so that time and distance are measured in units of the black 
hole mass, $M_{\rm BH}$. The GRMHD code uses the test-fluid approximation, so 
that energy of the orbiting fluid does not alter the background spacetime. We 
express mass and energy in relation to the maximum initial torus values. The 
simulation output consists of an extensive collection of data dumps, obtained at 
increments of $2\,M$ of simulation time; these dumps are used to compute 
shell-averaged fluxes to measure energy output and temporal variability (see DHK 
for details). 

\section{Overview of Disk/Jet Simulations}

The initial state of our
simulations is similar to that of DHK: a torus with a near-Keplerian 
rotation profile is seeded with a weak poloidal magnetic field (the strength of 
the field is set using the ratio of average gas to magnetic pressure in the 
torus, $\beta_{\rm disk}=\langle P_{\rm gas}\rangle/\langle P_{\rm mag}\rangle 
=$ 100) to trigger the MRI and generate an accretion flow. 
As noted above, two important code improvements allow us to probe the
dynamics of the axial funnel more thoroughly: an expanded dynamic range and
a significantly increased ceiling on the Lorentz factor (now at 50). These
improvements are crucial to addressing issues of jet dynamics.
Furthermore, to follow the evolution of the jets away from the inner engine,
we use a greatly extended radial range (in the 2D simulations). 
Also,
the grid outside the initial torus contains a radially infalling dust (the Bondi 
solution discussed in Hawley, Smarr \& Wilson, 1984; hereafter HSW) to provide 
an influx of matter against which the emerging jets must compete. 
The density contrast between dust and torus, at $10^{-6}$ (see Table 
\ref{params}), was chosen to be larger than the typical value
of a collapsar environment (W93) for reasons that will be
touched upon in the following sections.
In some 
simulations, this torus is also embedded in an external vertical magnetic field 
(Wald, 1974; hereafter W74) to 
gain insights into
the possible effects of such ambient fields. The strength of these fields was chosen so that they could potentially 
interact with the MRI-generated fields, as measured by $\beta_{\rm dust}$, the 
ratio of gas to magnetic pressure outside the initial torus. The choice of initial strength (see Table \ref{params}) was established in tests which 
revealed that strong fields tended to disrupt the initial torus through magnetic tension; ``intermediate'' values were chosen for these simulations.

The set of simulations consists of 6 models: the S (Schwarzschild, specific angular momentum $a/M=$ 0), R (Rapidly rotating Kerr hole, $a/M=$ 0.9), and E 
(Extreme Kerr hole, $a/M=$ 0.995) models,  with or without an external vertical 
field (distinguished by the subscript ``vf'', e.g., models E and ${\rm E}_{\rm 
vf}$). As shown in Table \ref{params}, the structural parameters of the initial 
tori place the initial pressure maximum close to the black hole; this 
approximates collapsar initial conditions (W93) and also provides a rapid 
evolution since the MRI which drives accretion operates on the orbital time 
scale of the initial torus, $T_{\rm orb}/M \approx$ 400. Simulations follow the 
evolution of the accretion disk through 5 orbits ($t_{\rm max}/M=$ 2000) for 
models R and E, and roughly 8 orbits for the S models. We also carried out one 
3D simulation analogous to model ${\rm R}_{\rm vf}$, though with a smaller 
radial extent. Such a simulation is important to shed light on possible 
artifacts due to axisymmetry.

The data collection interval, which begins at $t/M=$ 1200 for the high-spin
models ($t/M=$ 2400 for Schwarzschild) and extends for two full orbits of the main accretion disk. The 
start of the interval is chosen to avoid the early transient stage when the 
initial torus transforms into a turbulent accretion disk; the mass and energy 
fluxes during this transient phase would generate highly misleading results. 
After the transient phase, the accretion disk enters a quasi-steady state where 
the MRI-driven turbulence transports angular momentum outward, and accretion and 
ejection also proceed in a sustained manner. Extending the data gathering 
interval for two orbits ensures that representative time-averaged fluxes are 
obtained.

\section{Simulation Results}

The broad features of the simulations\footnote{See http://capca.ucalgary.ca/capca/ for animations.} are similar to those discussed in DHK 
and D05: the formation of a turbulent accretion flow after roughly two orbits of 
the main torus and the emergence of an evacuated funnel containing an unbound 
outflow driven by a largely radial magnetic field. An extensive unbound coronal 
wind occupies the space between the main accretion disk and the funnel outflow. 
We use specific binding energy, $e_{\rm bind}=-h\,U_t$, to distinguish the jet 
($e_{\rm bind} > 1.5$) from the unbound coronal outflow ($1.5 \ge e_{\rm bind}
\ge 1$) and the main disk ($e_{\rm bind} < 1$). Jets launched by a rapidly 
spinning black hole ($a/M=$ 0.9, 0.995) are energetic; those launched by a 
non-rotating hole are considerably weaker. The jets have a two-component 
structure: a hot, fast, tenuous outflow in the funnel proper, and a slower, 
colder, massive jet along the funnel wall. The funnel-wall jets are episodic in 
nature, fed by injection events where the accretion flow, corona, and funnel 
converge, at a radial distance comparable to the marginally stable orbit, 
$r_{\rm ms}$. 
The most significant difference between the S models and the high-spin models is 
that jet material in the S models is much slower to reach the outer radial boundary (hence the extended simulation time). Animations of gas density show 
that the 2D R- and E-model jets have displaced the initial radially infalling 
gas from the funnel by $t/M \approx$ 1200, whereas the S-model jets only do so 
by $t/M \approx$ 2400. 

\subsection{Jet Launching Mechanism}

The mechanism by which jets are launched was discussed in D05; the following
summary reiterates the salient points of the complex dynamics by which unbound 
outflows emerge from the accretion flow. In the region of the accretion flow 
near the marginally stable orbit (the inner torus region in the terminology of 
DHK), both pressure gradients and the Lorentz force act to lift material away from the equatorial plane. Some of this material is launched 
magneto-centrifugally in a manner reminiscent of the scenario of Blandford \& 
Payne (1982), generating the coronal wind; some of this material, which has too 
much angular momentum to penetrate the centrifugal barrier, also becomes part of 
the massive funnel-wall jet. There is also evidence that the low-angular 
momentum funnel outflow originates deeper in the accretion flow. Some of this 
material is produced in a gravitohydromagnetic (GHM) interaction in the 
ergosphere (Punsly \& Coroniti, 1990; Punsly, 2005), and possibly in a process 
similar to that proposed by Blandford \& Znajek (1977) where conditions in the 
ergosphere approach the force-free limit. The material in the funnel outflow is 
accelerated by a relatively strong, predominantly radial Lorentz force; gas 
pressure gradients in the funnel do not contribute significantly. There is also 
a significant amount of entrainment at the interface between the funnel wall and 
corona.

\subsection{Lorentz Factor and Enthalpy Knots}

\begin{figure}[ht]
    \epsscale{1.15}
    \plotone{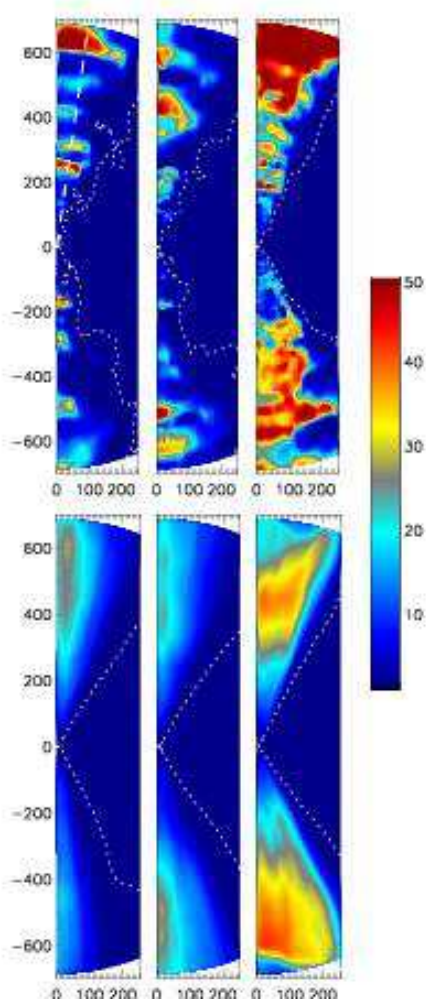} 
    \caption{\label{f1} 
    Lorentz factor, $W$, in the funnel outflow. In all panels, the 
    axes are in units of $M$ ($1\,M \approx 4$ km scaled to 
    collapsar dimensions). The black hole is located at the origin. 
    The top row, left to right, shows plots of 
    $W(r,\theta)$ for models ${\rm E}_{\rm vf}$ and ${\rm R}_{\rm vf}$
    at $t/M=$ 2000M, and ${\rm S}_{\rm vf}$ at $t/M=$ 3200. The 
    dotted contour corresponds to $e_{\rm bind}=1.5$ and marks the boundary of 
    the jets. The only region where elevated values 
    of $W$ are found is in the jets (and also in the bound plunging inflow 
    near the black hole, which is not resolved at the scale of this figure). 
    Maximum values of $W$ are found in knots that appear episodically in the 
    upper and lower parts of the funnel. The bottom three panels show, from left 
    to right, the time-averaged value of the Lorentz factor, 
    $\langle W(r,\theta) \rangle_t$, for models ${\rm E}_{\rm vf}$ and
    ${\rm R}_{\rm vf}$ ($1200\,M \le t \le 2000\,M$), and 
    ${\rm S}_{\rm vf}$ ($2400\,M \le t \le 3200\,M$). These plots show 
    evidence of an extended acceleration zone: 
    large Lorentz factors are built up over the full radial range. The plots 
    also show evidence of spin-dependent 
    collimation: for $r/M \gtrsim 300$, cylindrical collimation is seen in the
    high-spin models (E,R), while the zero-spin model shows no such
    collimation. The plots also show a model-dependent asymmetry; this feature 
    does not depend on the presence of an initial vertical field. The dashed 
    line in the top left panel is the line  
    of constant $\theta$ along which the cut shown in Figure \ref{f2} was taken. 
  } 
\end{figure}

Turning now to the question of Lorentz factor, $W$, in the funnel outflow, 
Figure \ref{f1} shows $W(r,\theta)$ for the ${\rm E}_{\rm vf}$, 
${\rm R}_{\rm vf}$, and ${\rm S}_{\rm vf}$ models at the end of the 
simulations, as well as time-averaged values, $\langle W(r,\theta) \rangle_t$ 
for each model. Elevated values of the Lorentz factor are found in compact, hot, 
evacuated knots that ascend the funnel radially; the combination of
low density and high temperature in the knots yields high specific
enthalpy, $h=1+\epsilon+P/\rho$\footnote{Ouyed et al. (1997) discuss the 
appearance of knots in MHD jet simulations; the knots seen in those
simulations consisted of high density material, in contrast to what is
observed in the present simulations.}. The highest values of Lorentz 
factor reach the maximum allowed by the code ($W_{\rm max} = 50$, and test runs 
show that much higher values could be reached with a higher ceiling [see 
Appendix]), and these are only found at large radii, suggesting the presence of 
an extensive region where the knots are gradually accelerated to higher Lorentz factors. The time-averaged plots contain three 
interesting features: i) maxima at large radii, ii) asymmetry, and iii) 
spin-dependent collimation. 

i) Time-averaging smoothes out the passage of the 
knots and reinforces the notion of an extensive acceleration zone; the Lorentz 
factor increases smoothly from the base of the jet, near $r_{\rm ms}$, outward 
to $r/M\approx$ 400---600, where the maxima are found. 

ii) The asymmetry in the north and south components of the
jets, which is seen both in the instantaneous and time-averaged plots, depends 
on the details of injection at the base of the jet and on the time interval 
for data collection (the dominance of one or the other funnel will reverse in the course of long simulations). 

iii) Perhaps the most striking feature of the time-averaged plots 
is the suggestion of strong cylindrical collimation in the high-spin models; 
whereas the S model shows a time-averaged pattern with a constant opening angle 
of approximately 16 degrees, the plots of the high-spin models indicate that the 
ultra-relativistic components of the jets run roughly parallel to the axis for 
$r/M \gtrsim$ 300. For the models with no initial vertical field, we find 
similar patterns for $W(r,\theta)$ and $\langle W(r,\theta) \rangle_t$ with one 
important exception: the maximum values seen in the time-averaged plots are 
systematically lower where no initial external field is present. Collimation is 
not enhanced by the initial external field in our simulations; as is the
case in the simulations with no initial external field, the properties
of the funnel magnetic field are largely determined by the accretion flow (D05).

\begin{figure}[ht]
    \epsscale{1.0}
    \plotone{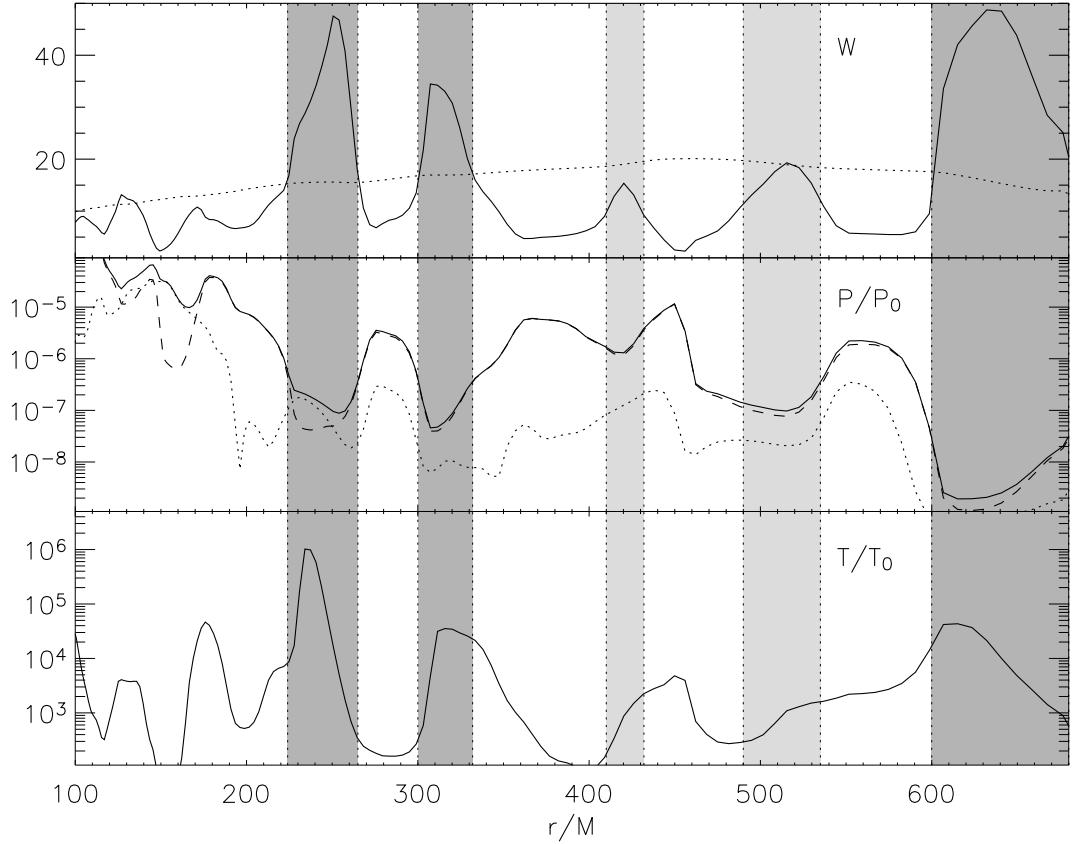} 
    \caption{\label{f2} 
    Radial cut through the upper funnel at $t=2000M$ for model 
    ${\rm E}_{\rm vf}$ ($a/M=0.995$;
    see Figure \ref{f1} for location of the cut). The top panel shows the 
    Lorentz factor, $W(r)$, at $t/M=$ 2000 (solid line), and $\langle W(r) 
    \rangle_t$ (dotted line). Knots with $W > 20$ are shaded in dark gray; knots 
    with $10 < W < 20$ are shaded lighter gray. The shading is extended to lower 
    panels to help align features in other quantities. The second panel from the 
    top shows pressure (total --- solid line, gas --- dashed, magnetic --- 
    dotted line) scaled to the maximum pressure in the initial torus. 
    The bottom panel shows temperature, scaled to the 
    maximum temperature in the initial torus; the knots with elevated $W$ 
    are $10^4$ to $10^6$ times hotter than the initial torus.} 
\end{figure}

The dashed line in the top left panel of Figure \ref{f1} indicates the angle at 
which a cut was taken through the knots for model ${\rm E}_{\rm vf}$. Data along 
this cut is shown in Figure \ref{f2}. The top panel shows $W(r)$ as a solid line 
and $\langle W(r)\rangle_{\rm t}$ as a dashed line. The grey shaded regions extending to the lower panels help locate the knots against other code variables. The second panel from the top shows pressure (total, gas, and 
magnetic), referenced to the initial torus maximum pressure; the knots sit in 
the troughs of outbound pressure waves. The bottom panel shows temperature, 
referenced to the initial torus maximum temperature; the knots, which are at 
least in part heated by shocks, are significantly hotter than the surrounding 
jet material, and considerably hotter than the initial torus. 

\subsection{Jet Variability}

\begin{figure}[ht]
    \epsscale{1.0}
    \plotone{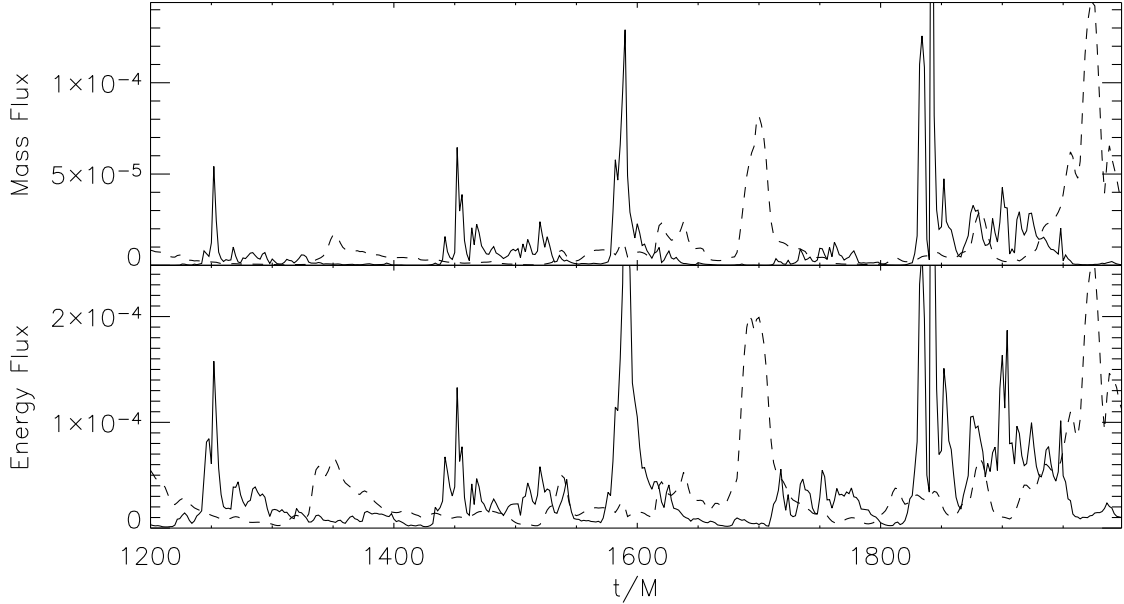} 
    \caption{\label{f3} 
    Shell-averaged mass and energy flux in the jets at 
    $r/M=15.0$ (solid line) and at $r/M=100$ (dashed line) for model 
    ${\rm E}_{\rm vf}$ ($a/M=0.995$). The horizontal axis is in units of M 
    (1 M $\approx 10^{-5}$ s, scaled to collapsar dimensions). The
    plots show that fluctuations occur on the order of the horizon-crossing 
    time near the black hole, and that bursts of elevated readings are followed 
    by extended quiescent periods, on the order 100 horizon-crossing times.
    At larger radii, the pattern of bursts and quiescent periods persists,
    but small-scale fluctuations are smoothed out by processes taking
    place higher in the funnel (e.g. entrainment, shocks). 
    The bursts in these figures correspond to the passage of knots through the 
    funnel. 
  } 
\end{figure}

The jets show a great deal of temporal variability. Figure \ref{f3} shows the 
time dependence of shell-averaged radial mass flux ($\dot{M}=\langle \rho\,U^r\rangle$) and energy flux ($\dot{E}=\langle{T^r}_t\rangle$, including
both the dominant fluid enthalpy component and the electromagnetic component) 
of jet material ($e_{\rm bind} > 1.5$) at $r/M = $ 15, near the base of the jet, 
and at $r/M =$ 100, in the acceleration zone for model ${\rm E}_{\rm vf}$. Close 
to the black hole, fine structure is detectable at a time scale comparable to 
the horizon light-crossing time. Energy flux comes in intense bursts 
corresponding to the passage of knots through the region, followed by extensive 
quiescent periods. At larger radii, variability follows a similar pattern, but 
the fine structure tends to be smoothed out by processes taking place higher in 
the funnel, such as entrainment (D05) and shocks. 

\subsection{Mass and Energy Flux}

Table \ref{results} summarizes the normalized rates of mass ($\dot{M}$) and 
energy ($\dot{E}$) transported by the jets and coronal wind; these quantities
are computed by integrating numerical fluxes through shells at $r/M=$ 100 (to 
compare 2D and 3D models) and normalized to initial torus mass ($M_0$) and 
energy ($E_0$), and hence are in units of inverse simulation time ($M^{-1}$).
It is important to emphasize that these calculations are done in a region where 
the funnel outflow is well established \emph{and} the numerical ceiling on the 
Lorentz factor does not come into play. 

To account for mass loss from the disk 
to the black hole, we compute the rate of mass accretion through the inner 
radial boundary. The mass and energy carried by the jets and coronal wind increase with increasing black hole spin, while the mass accreted onto the black hole decreases with increasing spin, consistent with the findings of D05. 
However, the energy in the jets is higher than in the analogous models in D05, 
in part due to initial conditions, but also due to improvements to the GRMHD 
code's ability to handle the dynamics of the jets. The mass and energy carried 
by the jets and coronal wind in 3D model ${\rm R_{vf}}$3D are comparable to the analogous 2D model ${\rm R_{vf}}$; this suggests that the other axisymmetric 
models are adequately capturing the fluxes of unbound mass and energy. The main 
difference between the 3D model and its 2D analog is that the jet is slightly 
more energetic than the coronal wind in 3D, while the situation is reversed in 
2D.

\section{Comparison to Collapsar Scenario}

To extract physical units from the mass and energy fluxes, we use the 
light-crossing time $t_{\rm lc}=G\,M_{\rm BH}/c^3$ and the gravitational length 
$r_g=G\,M_{\rm BH}/c^2$ to rescale time and distance measurements. We take 
collapsar fiducial parameters comparable to those of W93, namely a black hole 
mass of $M_{\rm BH} = 3\,M_\odot$ and an initial torus of mass $0.3\,M_\odot$, a 
value consistent with the initial torus parameters and a central density in the 
range of $10^{11}$---$10^{12}\,{\rm g}\,{\rm cm}^{-3}$. 

In 3D simulations, the lifetime of the jets is set by the reservoir of mass in 
the main disk; as long as turbulence drives the transport of angular momentum,
accretion and ejection will continue. In axisymmetry, the MRI which drives 
accretion is not sustainable, so a premature end to the accretion/ejection 
process occurs when the MRI subsides (after approximately 10 orbits of the main 
disk). Nevertheless, we use the mass fluxes ($\dot{M}$) of the axisymmetric 
models to infer the lifetime of the jets. This lifetime is determined by the 
depletion of disk mass, i.e. mass carried by the jet and coronal wind (through 
$r/M$=100 here) and the mass accreted into the black hole. Table \ref{results} 
lists the jet lifetimes inferred from the simulation data; the values range from 
0.1 to 0.2 s for the high-spin simulations to 1.0 to 1.4 s for the zero-spin 
case. Since the lifetime for model ${\rm R_{vf}}$3D is comparable to its 
axisymmetric analog, we can infer that jet lifetimes of a few tenths to a few 
seconds are predicted by these simulations, with longer lifetimes produced by the lower mass fluxes generated by low-spin black holes. 

The jet lifetime is used to calculate the energy transported by the jets through 
the simple relation $E_{\rm total} = t_{\rm eject}\,\dot{E}\,E_{\rm torus}$, 
where $E_{\rm torus} = 0.3\,M_\odot\,c^2$ is the rest mass of the initial torus. 
In all simulations, the energy carried by the jets is between  $10^{48}$ and 
$10^{49}$ erg
and some fraction of this energy should be converted to radiation once
jet material reaches the optically thin outer region of the collapsar.
An analogous calculation shows that the coronal wind carries a 
comparable amount of energy. These energetic coronal winds flow through a much 
larger surface area than the jets, but their physical effects on the outer 
layers of the remnant fallback stellar material are likely non-negligible. 
Our choice of density contrast between initial torus and infalling
dust reinforces this observation: the coronal winds had to reverse a 
substantial influx of material to become established.

\section{Summary and Discussion}

We have carried out a series of GRMHD simulations of accretion disk/jet systems 
in the spacetime of Kerr black holes. One of the most significant aspects of 
these simulations is that the entire accretion/ejection system is evolved in a 
self-consistent manner from a weakly magnetized initial torus. Our main 
motivation for this work was to determine what such simulations could say about 
the inner engine of the canonical collapsar model of $\gamma$-ray bursts (W93). 
GRBs are energetic explosions thought to originate during the 
collapse of massive stars; a significant fraction of the rest mass of the 
progenitor can be released during such bursts. Recent evidence (Sazonov et al., 
2004; Soderberg et al., 2004) suggests that the energy liberated in the form of 
$\gamma$-rays spans a broader range than previously thought, from $10^{47}$ to 
$10^{51}$ erg. The duration of the bursts also spans a considerable range, in 
two broad classes: short duration bursts, less than $\sim 2$ s, featuring a hard 
$\gamma$-ray spectrum, and longer duration bursts $\sim 20$ s, featuring a 
softer spectrum (e.g. M\'esz\'aros, 2002).

The simulations discussed here, which probe the basic dynamics of the 
disk/black-hole/jet interaction using full GRMHD, suggest that the collapsar 
scenario favours events at the low end of the distribution of long-duration
bursts, on the order of a few seconds. The 
lifetime of the jets is set by the depletion of disk mass by the jets and a 
significant coronal wind, as well as by accretion onto the black hole. With a 
inner engine consisting of a high-spin black hole, and an initial torus of a 
few tenths of a solar mass, the jets have a lifetime of a few tenths of a 
second. In addition, the jet material shows a great deal of 
temporal variability, associated with the passage of knots of shock-heated gas 
through the funnel. These knots are progressively accelerated to higher Lorentz 
factors as they ascend the funnel. Results suggest that the Lorentz factor 
saturates, in a time-averaged sense, to $W \approx$ 30 to 50 at a radius of $r/M 
\approx $ 500 to 600 (220 to 260 km scaled to collapsar dimensions); however, 
this saturation may be an artifact of the ceiling on the Lorentz factor 
($W_{\rm max} = 50$) in the GRMHD code, and true saturation may occur at greater 
radii. Fendt and Ouyed (2004), using a relativistic MHD wind model, suggest that 
Lorentz factors on the order of a few hundred to a few thousand are obtainable 
at $r/M \gtrsim$ 2000. 

In the simulations, the energy transported by the jets is approximately 
$10^{49}$ erg. Since this energy is channeled into ultra-relativistic axial 
outflows (cylindrically collimated when driven by a high-spin black hole), strong beaming effects are to be expected. Given that the flux of mass and 
energy in the jets are good indicators of temporal variability in observables, 
the time-dependence of the fluxes suggests that emissions from the jets should 
show bursts of activity on the order of a few milliseconds, with slightly longer 
quiescent periods.  Our simulations indicate that the presence of an initial 
external (vertical) field can enhance some aspects of the jets, such as the 
maximum time-averaged Lorentz factor in the jets, while leaving other quantities 
essentially unchanged; in general terms, the magnetic field generated by the MRI 
is the main driver of the accretion flow and hence the jets, which are fed by 
magnetized fluid injected at their base from the accretion flow. The time-
averaged distribution of Lorentz factor shows a spin-dependent collimation of 
the ultra-relativistic jet material; in the high-spin simulations, the jets seem 
to become cylindrically collimated for $r/M \gtrsim$ 300 (130 km scaled to 
collapsar dimensions).

Based on our simulations, the action of the inner engine in the canonical 
collapsar model of W93 can only power GRBs at the low end of the distribution 
of long-duration bursts. However, the collapsar model also envisages feeding the 
accretion disk over time from the outer layer of a Wolf-Rayet star. Our results show that such a process would be disrupted following the establishment of the 
strong coronal winds (which can easily displace a considerable amount of 
infalling dust). Assuming that some remnant material does manage to make its way 
to the main accretion disk, then the lifetime and total energy of the inner 
engine could be plausibly extended. However, this finding does not necessarily 
argue against the collapsar model in explaining extreme duration bursts; as 
discussed by Toma et al. (2005), the GRBs at the upper end of the distribution 
for long-duration bursts can be explained as statistical outliers related to 
observational effects.

It is important to emphasize that these simulations probe only the dynamics of 
the collapsar model in a fixed background spacetime. More sophisticated 
treatments should ideally incorporate a dynamical black hole spacetime 
(especially if more massive accretion disks are to be modeled) as well as self-
consistent calculations of radiative transport. Such a code is not yet 
available; however, it is hoped that some progress can be made at extracting 
observables from the fluxes reported here by coupling the GRMHD code to a ray 
tracer and a suitable emission model, which will provide a more complete picture 
of the time-dependent signals received by distant observers. This work is 
currently underway, and should be reported in the near future.

\appendix{\bf Appendix: Equations of GRMHD; Lorentz Factor Calculations}

The GRMHD code, described in detail in DH03, is used to study the dynamical 
properties of a magnetized 
fluid in the background spacetime of a Kerr black hole by numerically 
solving the equation of continuity, $\nabla_\mu \left(\rho\,U^\mu\right)=0$,
the energy-momentum conservation law, $\nabla_\mu\,T^{\mu \nu} = 0$,
and the Maxwell's equations, 
$\nabla_\mu F^{\mu \nu} = 4\,\pi\,J^\nu$ and 
$\nabla_\mu {}^*F^{\mu \nu} = 0$, which, in the MHD approximation,
reduce to the induction equation
$\partial_\delta\,F_{\alpha \beta} + 
\partial_\alpha\,F_{\beta \delta} + 
\partial_\beta\,F_{\delta \alpha} = 0$. 
The GRMHD code evolves a set of primitive and secondary code variables
directly; these variables were chosen to correspond directly, or 
through simple relations, to physical variables (e.g. magnetic field, gas
density, velocity). This avoids costly
calculations to extract physical variables from tensor quantities.
In the above expressions,
$\rho$ is the density, $U^\mu$ the 4-velocity, 
$T^{\mu \nu}$ the energy-momentum tensor, and
$F_{\alpha \beta}$ the electromagnetic field strength tensor.
The energy momentum tensor is given by 
$T^{\mu \nu} = \left[ \left(\rho\,h+{\|b\|}^2\right)\,U^\mu\,U^\nu + 
\left(P+{{\|b\|}^2 \over 2}\right)\,g^{\mu \nu} - b^\mu\,b^\nu\right]$
where $h=1 + \epsilon + P/\rho$ is the specific 
enthalpy, with $\epsilon$ the specific internal energy and 
$P=\rho\,\epsilon\,(\Gamma-1)$ the ideal gas pressure ($\Gamma$ is the
adiabatic exponent); ${\|b\|}^2=b^\mu\,b_\mu$ is the magnetic field 
intensity; and $b^\mu = {}^*F^{\mu \nu}\,U_\nu/(4\,\pi)$ is the
magnetic field 4-vector. 
The induction equation is rewritten in terms of the Constrained Transport
(CT; Evans \& Hawley, 1988) magnetic field variables 
${\cal B}^i = \epsilon_{i j k}\,F_{j k}$,
as $\partial_t {\cal B}^i - 
\partial_j\left( V^i\,{\cal B}^j - V^j\,{\cal B}^i\right)= 0$,
where $V^i=U^i/U^t$ is the transport velocity, and $U^t=W/\alpha$,
with $W$ the Lorentz factor. The CT algorithm ensures that the
constraint $\partial_i {\cal B}^i = 0$ is satisfied to rounding
error. 
The Kerr metric is expressed in Boyer-Lindquist coordinates, for
which ${ds}^2=g_{t t}\,{dt}^2+2\,g_{t \phi}\,{dt}\,{d \phi}+g_{r r}\,{dr}^2 +
g_{\theta \theta}\,{d \theta}^2 +g_{\phi \phi}\,{d \phi}^2$;
$\alpha = {(-g^{tt})}^{-1/2}$ is the lapse function.

The 4-velocity is subject to the constraint $U^\mu\,U_\mu=-1$. Defining the 
4-momentum as $S_\mu = (\rho\,h\ + {\|b\|}^2)\,W\,U_\mu$, we obtain the equivalent condition $S^\mu S_\mu= -{(\rho\,h\,W + {\|b\|}^2\,W)}^2$.
In the GRMHD code, the fundamental variables $D=\rho\,W$ and $E=D\,\epsilon$ are 
used to write $\rho\,h\,W=D+\Gamma\,E$; the spatial components of the 
4-momentum are treated as fundamental variables and evolved using the momentum 
equations (equation (30) of DH03); and the Lorentz factor, $W$, is 
extracted using the normalization condition at the end of each time step (along 
with $V^i$). It is straightforward to show that the normalization condition 
yields the following quartic expression for $W$,
\begin{equation}\label{wquartic}
\eta^2 + {(1 + \xi\,W)}^2\,(1-W^2) = 0
\end{equation}
where $\eta = {\|\tilde{S}\|}/(D+\Gamma\,E)$, 
$\xi = {\|b\|}^2/(D+\Gamma\,E)$, and 
${\|\tilde{S}\|}^2=S_r^2/g_{rr}+S_\theta^2/g_{\theta \theta}+S_\phi^2/
g_{\phi \phi}$. 
In this approach, the solutions for the Lorentz factor span a two-dimensional
parameter space, $W(\eta,\xi)$, with $\eta$ analogous to a fluid velocity and
$\xi$ analogous to an Alfv\'en velocity. Using the analytic expression
for the physically allowed root of equation (\ref{wquartic}) provides a 
significant improvement over the simpler (and faster) calculation used in earlier 
versions of the GRMHD code. Though costlier from a numerical point of view, this 
quartic solution significantly raises the maximum allowable value of the Lorentz 
factor from $W_{\rm max}=$ 8 in the original code to 50 in the current version. 
This upper bound, which occurs in the limit of $\xi/\eta \gg 1$, is set by 
rounding error (in double-precision arithmetic $W_{\rm safe} \lesssim 80$, 
but the more conservative limit of $W_{\rm max}=50$ has been used in the
simulations reported here).

\acknowledgements{
This research is supported by the Natural Sciences and Engineering Research
Council of Canada (NSERC), as well as the Alberta Ingenuity Fund (AIF).
We thank John Hawley for insightful comments on an early draft of this paper, 
and Martin Siegert of WestGrid for guidance on numerical issues regarding
quartic roots. The 3D numerical simulation which complements
the axisymmetric simulations was carried out on the SGI system Arcturus, a
WestGrid facility. WestGrid computing resources are funded in part by the 
Canada Foundation for Innovation, Alberta Innovation and Science, BC Advanced Education, and the participating research institutions.}


\clearpage

\begin{deluxetable}{lllll}
\tablecolumns{5}
\tablewidth{0pc}
\tablecaption{Model-specific Initial Conditions and Simulation Parameters\label{params}}
\tablehead{\multicolumn{2}{l}{Parameter} & \colhead{Schwarzschild (S)} & 
\colhead{Rapid Kerr (R)}& \colhead{Extreme Kerr (E)} } 
\startdata
Black hole: \\
\quad spin & $a/M$   & 0.000 & 0.900 & 0.995\\
\quad marg. stable orbit & $r_{\rm ms}/M$ & 6.00 & 2.32 & 1.34  \\

\tableline
Torus:\tablenotemark{\bf a}\\
\quad inner edge & $r_{\rm in}/M$  & 9.5 & 9.5 & 9.5\\
\quad pressure max.& $r_{\rm {P}_{\rm max}}/M$ 
                & 16.9 & 16.1 & 16.1 \\
\quad outer edge & $r_{\rm out}/M$  & 31 & 36 & 38\\
\quad maximum height & $H/M$  & 5 & 9 & 10\\
\quad orb. period (at $r_{\rm {P}_{\rm max}}$) & $T_{\rm orb}/M$ & 436 & 422 & 424\\
\quad mass & $M_0$ & 0.225 & 5.32 & 8.26\\
\quad rest energy & $E_0$ & 0.219 &  5.19 & 8.08\\
\quad ang. momentum & $L_0$ &  1.07  & 24.6 & 38.5\\

\tableline
Background:\tablenotemark{\bf b}\\
\quad density contrast & $\bar{\rho}_{\rm dust}/\bar{\rho}_{\rm disk}$
           & $10^{-6}$ & $10^{-6}$ & $10^{-6}$\\
\quad energy contrast & $\bar{\epsilon}_{\rm dust}/\bar{\epsilon}_{\rm disk}$
           & $10^{-11}$ & $10^{-11}$ & $10^{-11}$\\
\quad vertical field\tablenotemark{\bf c} & $B_0$ & $10^{-6}$ & $10^{-6}$ & $10^{-8}$ \\
               & $\beta_{\rm dust}$ & 
               $\sim 10^3-10^4$ & $\sim 10^2-10^3$ &$\sim 10^4-10^5$ \\
               
\tableline
Grid:\tablenotemark{\bf d}\\
\quad Inner boundary & $r_{\rm min}/M$ & 2.10 & 1.45 & 1.25 \\
\quad End time & $t_{\rm max}/M$ & 4000 & 2000 & 2000 \\
\enddata
\tablenotetext{a}{Equations given in DHK; as in DHK, $K=0.01$ and $q=1.68$. 
Equation of state uses $\Gamma=$ 4/3.
Torus seeded with poloidal magnetic field loops with 
$\beta=\langle  P_{\rm gas}\rangle/\langle  P_{\rm mag}\rangle \approx 100$. 
Torus also given small (1 \%) density perturbation.} 
\tablenotetext{b}{Equations given in HSW;
scaled to grid-averaged density and energy contrasts shown.}
\tablenotetext{c}{Field intensity set by constant $B_0$ (W74). Values of $\beta_{\rm dust}$, the ratio of gas pressure in the dust to vertical field magnetic pressure, are taken near the surface of the initial torus.}
\tablenotetext{d}{All 2D grids: $512^2$ zones; radial grid uses $\cosh$-scaling, $r_{\rm max}/M=700$; $\theta$-grid uses linear scaling with polar axis offset 
$\Delta \theta/\pi = 10^{-5}$. 3D grid: $192^3$ zones; radial $\cosh$-scaling 
with $r_{\rm max}/M=120$; $\theta$-grid exponentially scaled with $\Delta 
\theta/\pi = 10^{-3}$; $\phi$-grid spans linearly 0 to $2\,\pi$.}
\end{deluxetable}

\clearpage

\begin{deluxetable}{lccccccc}
\tablecolumns{8}
\tablewidth{0pc}
\tablecaption{Normalized Mass and Energy Fluxes\label{results}}
\tablehead{\colhead{Model} & \colhead{S}& \colhead{${\rm S}_{\rm vf}$} & \colhead{R}
& \colhead{${\rm R}_{\rm vf}$}& \colhead{${\rm R}_{\rm vf}$3D}& \colhead{E}& \colhead{${\rm E}_{\rm vf}$} } 
\startdata
Jets (Funnel Outflow): \\
\quad $\dot{M}\left(\times {10}^{-5}\right)$ 
              & 0.010 & 0.009 & 0.653 & 0.836 & 1.302 & 1.767 & 1.150 \\
\quad $\dot{E}\left(\times {10}^{-5}\right)$ 
              & 0.106 & 0.091 & 2.880 & 3.003 & 4.131 & 5.906 & 3.478 \\
\tableline
Coronal Wind: \\
\quad $\dot{M}\left(\times {10}^{-5}\right)$ 
              & 0.041 & 0.044 & 6.093 & 5.108 & 2.840 & 11.430 & 7.265 \\
\quad $\dot{E}\left(\times {10}^{-5}\right)$
              & 0.045 & 0.049 & 6.896 & 5.843 & 3.243 & 13.138 & 8.387 \\

\tableline
Black hole accretion: \\
\quad $\dot{M}\left(\times {10}^{-5}\right)$
              & 1.111 & 0.825 & 0.011 & 0.026 & 0.064 & 0.008 & 0.012 \\

\tableline
Cummulative:\\
\quad $t_{\rm eject}$ (s)& 
      1.084 & 1.380 & 0.214 & 0.237 & 0.316 & 0.109 & 0.170 \\
\quad $E_{\rm jet} \left(\times {10}^{48}\,{\rm erg}\right)$ & 
      0.615 & 0.782 & 3.298 & 3.804 & 6.961 & 3.440 & 3.162  \\
\quad $E_{\rm wind} \left(\times {10}^{48}\,{\rm erg}\right)$ & 
      0.262 & 0.334 & 7.896 & 7.400 & 5.465 & 7.652 & 7.626  \\
\enddata
\tablecomments{
Fluxes computed at $r/M=100$ (jet/wind) and at $r = r_{\rm min}$ (accreted mass).
Jet component satisfies $e_{\rm bind} > 1.5$; coronal wind satisfies
$1.5 \ge e_{\rm bind} \ge 1.0$. Fluxes are expressed as fractions of 
initial torus mass and energy: 
$\dot{M} = \langle \rho\,U^r\rangle/\left(M_0\,\Delta t\right)$ and
$\dot{E} = \langle -{T^r}_t\rangle/\left(E_0\,\Delta t\right)$, where 
the flux consists of both the dominant fluid enthalpy component and
the electromagnetic component). 
Angle brackets denote integration $\langle {\cal F}\rangle = 
\int\int\int{{\cal F}\,\sqrt{-g}\, d \theta\,d \phi\,dt}$; $\Delta t$ denotes
the interval of integration (1200 to 2000 M for 2D (R, E) models, 
2400 to 3200 M for S models, 1200 to 1700 M for ${\rm R}_{\rm vf}$3D).
Ejection time is computed using $t_{\rm eject}= t_{\rm M}/\dot{M}_{\rm tot}$ where
$\dot{M}_{\rm tot} \equiv \dot{M}_{\rm (r/M=100)}+ \dot{M}_{\rm (r=r_{min})}$ represents total mass loss through jets and coronal wind at $r/M=$ 100 and
accretion onto the black hole, converted to seconds using the light-crossing 
time $t_{\rm M}=G\,M\,c^{-3}$. The total energy is computed using 
$E_{\rm jet/wind} = t_{\rm eject}\,\dot{E}_{\rm jet/wind}\,E_{\rm torus}$
where $E_{\rm torus}=0.3\,M_\odot\,c^2$ is the initial torus rest mass in erg.
}
\end{deluxetable}
 
\clearpage

\clearpage

\clearpage

\end{document}